\documentclass[prd,aps,showpacs,nofootinbib,preprint,tightenlines]{revtex4}

\newcommand{\checked}[1]{}

\usepackage{graphicx}

\newcommand{\beq}{\begin{equation}}
\newcommand{\eeq}{\end{equation}}
\newcommand{\bqa}{\begin{eqnarray}}
\newcommand{\eqa}{\end{eqnarray}}

\begin{document}

\title{Collective Modes of an Anisotropic Quark-Gluon Plasma}

\preprint{ TUW-03-09 }

\author{Paul Romatschke}
\author{Michael Strickland}
\affiliation{Institut f\"ur Theoretische Physik, Technische Universit\"at Wien,
	Wiedner Hauptstrasse 8-10, A-1040 Vienna, Austria
     \vspace{1cm}
	}

\begin{abstract}
We analyze the collective modes of high-temperature QCD in the case when there
is an anisotropy in the momentum-space distribution function for the gluons.  We 
perform a tensor decomposition of the gluon self-energy and solve the dispersion
relations for both stable and unstable modes.  Results are presented for a class 
of anisotropic distribution functions which can be obtained by stretching or
squeezing an isotropic distribution function along one direction in momentum space.  
We find that there are three stable modes and either one or two unstable modes depending
on whether the distribution function is stretched or squeezed.  The presence of unstable modes
which have exponential growth can lead to a more rapid thermalization and 
isotropization of the soft modes in a quark gluon plasma and therefore may play 
an important role in the dynamical evolution of a quark-gluon plasma.  
\end{abstract}
\maketitle
\newpage

\small

\section{Introduction}

In the ongoing ultra-relativistic heavy collision experiments at the Relativistic Heavy Ion Collider
(RHIC) and the upcoming ones at the Large Hadron Collider (LHC) the behavior of nuclear matter 
under extreme conditions will be studied.  The hope of these experiments is to create 
temperatures which are high enough for nuclear matter to undergo a phase transition to 
a quark gluon plasma (QGP).  The quark gluon plasma, if generated, is expected to expand, cool, 
and then hadronize in the final stage of its evolution.  In this context, an outstanding question 
faced by experimentalists and theorists is whether or not the system will ``thermalize'' fast enough 
to allow a thermodynamic description of the system during the central part of its evolution.

In this paper we study the role of the collective modes of finite-temperature QCD in the 
thermalization, particularly the isotropization, of a finite-temperature QGP 
with anisotropic momentum-space distribution functions.  This question has been addressed in
previous papers in which the existence of instabilities of a QGP were studied.  In 
Refs.~\cite{SM:1993,SM:1994,SM:1997} Mr\'owczy\'nski discussed the existence of instabilities
to chromomagnetic fluctuations with a particular orientation of the chromoelectric field and 
wave-vector.  In those papers Mr\'owczy\'nski showed that there existed an instability 
which was the equivalent of the Weibel or filamentation instability in electrodynamics \cite{Weibel:1959}.  
Weibel showed in his original paper that, within electrodynamics, unstable transverse modes 
exist in plasmas with anisotropic momentum distributions and he also derived their rate of 
growth in linear response theory.  These types of instabilities are potentially very important to
QGP evolution at RHIC or LHC due to the large amount of momentum-space anisotropy in the 
gluon distribution functions at $\tau \sim 1\;{\rm fm/c}$.  

Mr\'owczy\'nski and Randrup have recently performed phenomenological estimates of the growth
rate of the instabilities for two types of anisotropic distribution functions \cite{RM:2003}.  
They found that the degree of amplification of the Weibel instability is not expected to dominate the
dynamics of a QGP instead being comparable to the contribution from elastic Boltzmann collisions.  
However, they did point out that since a large number of the unstable modes could be excited 
then it is possible that their combined effect on the overall dynamics could be significant.
In this paper we perform a detailed study of the hard-thermal-loop resummed gluon self-energy
including a complete tensor decomposition of the self-energy and identification of all stable
and unstable collective modes.

In Sec.~\ref{htl-sec} we set up the framework used to obtain the hard-thermal-loop 
self-energy in a system with an anisotropic momentum 
space distribution.  In Sec.~\ref{tensor-sec} we present a tensor decomposition of the 
self-energy and dielectric tensors.  In 
Sec.~\ref{struc-sec} we work out the details of the tensor decomposition and give 
expressions for the self-energy ``structure functions.''  In Sec.~\ref{static-sec} we 
discuss the static limit of the various self-energy structure functions.  In 
Sec.~\ref{mode-sec} we use the tensor decomposition of the dielectric tensor to determine 
dispersion relations for all stable and unstable modes. 
In Sec.~\ref{smallxi-sec} we present analytic expressions for the self-energy structure
functions in the small-anisotropy limit.
Finally, in Sec.~\ref{conc-sec} we present 
conclusions and outlook for application of the results found here.  We provide a summary
of our notational conventions and expressions for the various self-energy structure
functions in two appendices.

\section{Hard-Thermal-Loop Self-Energy}
\label{htl-sec}

We begin by repeating some of the steps necessary to derive the hard-thermal-loop resummed
gluon self-energy within semi-classical transport theory \cite{SM:1993,SM:1994,SM:1997}.  
Within this approach partons are described by their phase-space densities and their time 
evolution is given by Vlasov-type transport equations \cite{EH:1989,BI:2002}.  In this paper we
will concentrate on the physics at the soft scale, $k \sim g T \ll T$, which is the first scale
at which collective motion appears.  At this scale the magnitude of the field fluctuations
is $A \sim \sqrt{g} T$ and derivatives are of the scale $\partial_x \sim g T$.  With this power-counting 
a systematic truncation of the terms contributing to the transport equations for soft momenta 
can be realized.

At leading order in the coupling constant the color current, $J^\mu$, induced by a soft gauge field, 
$A^\mu$, with four-momentum $K=(\omega,{\bf k})$ can be obtained by performing a covariant 
gradient expansion of the quark and gluon Wigner functions in mean-field approximation.
The result is
\beq
J^{\mu,a}_{\rm ind}(X)=g \int \frac{d^3 p}{(2\pi)^3} V^{\mu} (2 N_c \delta N^a(p,X)+
	N_f ( \delta n_{+}^{a}(p,X)-\delta n_{-}^{a}(p,X)) \; ,
\label{current}
\eeq
\checked{mp}
where $V^{\mu} = (1,{\bf k}/\omega)$ is the gauge field four-velocity, $\delta N^a(p,X)$ is the 
fluctuating part of the gluon density, and $\delta n^a_{+}(p,X)$ and $\delta n^a_{-}(p,X)$ 
are the fluctuating parts of the quark and anti-quark densities, respectively.  Note that 
$\delta N^a$ transforms as a vector in the adjoint representation 
($\delta N \equiv \delta N^a T^a$) and $\delta n^a_{\pm}$ transforms as a vector
in the fundamental representation ($\delta n_{\pm} \equiv \delta n^a_{\pm} t^a$).

The quark and gluon density matrices above satisfy the following transport equations
\bqa
\label{qvlasov}
\left[ V \cdot D_{X}, \delta n_{\pm}(p,X) \right] &=& 
  \mp g V_{\mu} F^{\mu \nu}(p,X) \partial_{\nu} n_{\pm}({\bf p}) \; ,\\
\left[ V \cdot D_{X}, \delta N(p,X) \right] &=& 
  - g V_{\mu} F^{\mu \nu}(p,X) \partial_{\nu} N({\bf p}) \; ,
\label{gvlasov}
\eqa
\checked{mp}
where $D_X = \partial_X + i g A(X)$ is the covariant derivative.

Solving the transport equations (\ref{qvlasov}) and (\ref{gvlasov}) for the fluctuations 
$\delta N$ and  $\delta n_\pm$ gives the induced current via Eq.~(\ref{current})
\bqa
J^{\mu}_{\rm ind}(X)=g^2 \int \frac{d^3 p}{(2\pi)^3} V^{\mu} V^{\alpha} 
	\partial^{\beta}_{(p)} f({\bf p}) \int d\tau \, 
	U(X,X-V\tau) F_{\alpha \beta}(X-V\tau) U(X-V\tau,X) \; ,
\eqa
where $U(X,Y)$ is a gauge parallel transporter defined by the path-ordered integral
\beq
U(X,Y) = \mathcal{P} \, {\rm exp}\left[ - i g \int_X^Y d Z_\mu A^\mu(Z) \right] \; ,
\eeq
$F_{\alpha \beta} = \partial_\alpha A_\beta - \partial_\beta A_\alpha - i g [A_\mu,A_\nu]$ 
is the gluon field strength tensor, and
\beq
f({\bf p}) = 2 N_c N({\bf p})+ N_{f} (n_{+}({\bf p})+n_{-}({\bf p})) \; .
\eeq
Neglecting terms of subleading order in $g$ (implying $U\rightarrow1$ and $F_{\alpha \beta} 
\rightarrow \partial_\alpha A_\beta - \partial_\beta A_\alpha$) and performing a Fourier 
transform of the induced current to momentum space we obtain
\beq
J^{\mu}_{\rm ind}(K)=g^2 \int \frac{d^3 p}{(2\pi)^3} V^{\mu} %
\partial^{\beta}_{(p)} f({\bf p}) \left( g_{\gamma \beta} - %
\frac{V_{\gamma} K_{\beta}}{K\cdot V + i \epsilon}\right) A^{\gamma}(K) \; ,
\eeq
where $\epsilon$ is a small parameter that has to be sent to zero in the end.

From this expression of the induced current the self-energy is obtained via
\beq
\Pi^{\mu \nu}(K)=\frac{\delta J^{\mu}_{\rm ind}(K)}{\delta A_{\nu}(K)} \; ,
\eeq
which gives
\beq
\Pi^{\mu \nu}(K)= g^2 \int \frac{d^3 p}{(2\pi)^3} V^{\mu} %
\partial^{\beta}_{(p)} f({\bf p}) \left( g_{\nu \beta} - %
\frac{V_{\nu} K_{\beta}}{K\cdot V + i \epsilon}\right) \; .
\label{selfenergy1}
\eeq
This tensor is symmetric, $\Pi^{\mu\nu}(K)=\Pi^{\nu\mu}(K)$, and transverse, 
$K^\mu\Pi^{\mu\nu}(K)=0$.  Note that the same result can be obtained 
using diagrammatic methods if one assumes that the distribution function is symmetric under 
${\bf p} \rightarrow - {\bf p}$~~\cite{MT:2000}.  

In the linear approximation the equation of motion for the gauge fields can be obtained
by expressing the induced current in terms of the self-energy
\beq
J^\mu_{\rm ind}(K) = \Pi^{\mu\nu}(K) A^\nu(K) \; ,
\eeq
and plugging this into Maxwell's equation
\beq
-iK_\mu F^{\mu\nu}(K) = J^\nu_{\rm ind}(K) + J_{\rm ext}^\nu(K) \; ,
\eeq
to obtain
\beq
[K^2 g^{\mu\nu} - K^\mu K^\nu + \Pi^{\mu\nu}(K)]A_\nu(K) = - J_{\rm ext}^\nu(K) \; ,
\eeq
where $J^\nu_{\rm ext}$ is an external current.
Using the gauge invariance of the self-energy we can write this 
in terms of a physical electric field by specifying a particular
gauge.  In the temporal axial gauge defined by $A_0=0$ we obtain 
\beq
[(k^2-\omega^2)\delta^{ij} - k^i k^j + \Pi^{ij}(K)] E^j(K) = 
  (\Delta^{-1}(K))^{ij} E^j(K) = i \omega \, J_{\rm ext}^i(K) \; .
\eeq
Inverting the propagator allows us to determine the response of
the system to the external source
\beq
E^i(K) = i \omega \, \Delta^{ij}(K) J_{\rm ext}^j(K) \; .
\eeq
The dispersion relations for the collective modes can be obtained by
finding the poles in the propagator $\Delta^{ij}(K)$.

\section{Tensor Decomposition}
\label{tensor-sec}

In this section we develop a tensor basis for an anisotropic system in
which there is only one preferred direction.
As mentioned above the self-energy is symmetric and transverse. As a result 
not all components of $\Pi^{\mu \nu}$ are 
independent and we can restrict our considerations
to the spatial part of $\Pi^{\mu \nu}$, denoted $\Pi^{i j}$.
We therefore need to construct a basis for a symmetric 3-tensor that
-- apart from the momentum $k^{i}$ -- also depends on a 
fixed anisotropy three-vector $n^{i}$, with $n^2=1$.
 Following Ref.~\cite{KKR:1991} we first define
the projection operator
\begin{equation}
A^{ij}=\delta^{ij}-k^{i}k^{j}/k^2,
\end{equation}
\checked{mp}
and use it to construct $\tilde{n}^{i}=A^{ij} n^{j}$ which obeys 
$\tilde{n} \cdot k =0$. With this we can construct the remaining
three tensors
\begin{equation}
B^{ij}=k^{i}k^{j}/k^2
\end{equation}
\begin{equation}
C^{ij}=\tilde{n}^{i} \tilde{n}^{j} / \tilde{n}^2
\end{equation}
\begin{equation}
D^{ij}=k^{i}\tilde{n}^{j}+k^{j}\tilde{n}^{i}.
\end{equation}
\checked{mp}
Any symmetric 3-tensor ${\bf T}$ can now be decomposed into the basis spanned by 
the four tensors ${\bf A},{\bf B},{\bf C},$ and ${\bf D}$
\beq
{\bf T}=a\,{\bf A}+b\,{\bf B}+c\,{\bf C}+d\,{\bf D} \; .
\eeq
\checked{mp}
Furthermore, the inverse of any such tensor is then given as
\beq
{\bf T}^{-1}=a^{-1}{\bf A}+\frac{(a+c){\bf B}-a^{-1}(bc-\tilde n^2 k^2 d^2 ){\bf C}-d{\bf D}}{
b(a+c)- \tilde n^2 k^2 d^2 } \; .
\label{inversion}
\eeq
\checked{mp}

\section{Self-energy Structure Functions}
\label{struc-sec}

The spacelike components of the self-energy tensor can be written as
\begin{equation}
\Pi^{i j}(K) = - g^2 \int \frac{d^3 p}{(2\pi)^3} v^{i} \partial^{l} f({\bf p})
\left( \delta^{j l}+\frac{v^{j} k^{l}}{K\cdot V + i \epsilon}\right) \; .
\label{selfenergy2}
\end{equation}
\checked{mp}
At this point the distribution function $f({\bf p})$ is completely arbitrary.
In order to proceed we need to specify a form for the
distribution function.  In what follows we will assume that
$f({\bf p})$ can be obtained from an arbitrary isotropic distribution 
function by the rescaling of only one direction in momentum space.  In practice 
this means that, given any isotropic distribution function 
$f_{\rm iso}({\bf p}^2)$, we can construct an anisotropic version
by changing the argument 
\begin{equation}
f({\bf p}) = f_{\rm iso}\left({\bf p}^2+\xi({\bf p}\cdot{\bf \hat n})^2\right) \; ,
\end{equation}
\checked{mp}
where ${\bf \hat n}$ is the direction of the anisotropy and $\xi>-1$ is a adjustable
anisotropy parameter.  Note that $\xi>0$ corresponds to a
contraction of the distribution in the ${\bf \hat n}$ direction whereas $-1<\xi<0$ 
corresponds to a stretching of the distribution in the ${\bf \hat n}$ direction. 
This assumption allows us to simplify (\ref{selfenergy2}) by performing a change of
variables to $\tilde p$
\begin{equation}
\tilde{p}^2=p^2\left(1+\xi ({\bf v}\cdot{\bf n})^2\right) \; .
\end{equation}
\checked{mp}
After making this change of variables it is possible to integrate out the 
$|\tilde p|$-dependence giving
\begin{equation}
\Pi^{i j}(K) = m_{D}^2 \int \frac{d \Omega}{4 \pi} v^{i}%
\frac{v^{l}+\xi({\bf v}.{\bf n}) n^{l}}{%
(1+\xi({\bf v}.{\bf n})^2)^2}
\left( \delta^{j l}+\frac{v^{j} k^{l}}{K\cdot V + i \epsilon}\right) \; ,
\end{equation}
\checked{mp}
where 
\beq
m_D^2 = -{g^2\over 2\pi^2} \int_0^\infty d p \,  
  p^2 {d f_{\rm iso}(p^2) \over dp} \; .
\eeq
\checked{mp}
We can then decompose the self-energy into four structure functions
\begin{equation}
\Pi^{ij}=\alpha A^{i j}+\beta B^{ij} + \gamma C^{ij} + \delta D^{ij} \; ,
\end{equation}
\checked{mp}
which are determined by taking the following contractions:
\begin{eqnarray}
k^{i} \Pi^{ij} k^{j} & = & k^2 \beta \; , \nonumber \\
\tilde{n}^{i} \Pi^{ij} k^{j} & = & \tilde{n}^2 k^2 \delta \; , \nonumber \\
\tilde{n}^{i} \Pi^{ij} \tilde{n}^{j} & = & \tilde{n}^2 (\alpha+\gamma) \; , \nonumber \\
{\rm Tr}\,{\Pi^{ij}} & = & 2\alpha +\beta +\gamma \; .
\label{contractions}
\end{eqnarray}
\checked{mp}
In Appendix \ref{exp-app} we collect the resulting integral expressions
for the structure functions.
All four structure functions depend on $m_D$, $\omega$, $k$, $\xi$, and
${\bf \hat k}\cdot{\bf \hat n}=\cos\theta_n$.
In the limit $\xi \rightarrow 0$ the structure functions $\alpha$ and $\beta$ 
reduce to the isotropic hard-thermal-loop self-energies and $\gamma$ and $\delta$
vanish
\begin{figure}
\includegraphics[width=14.5cm]{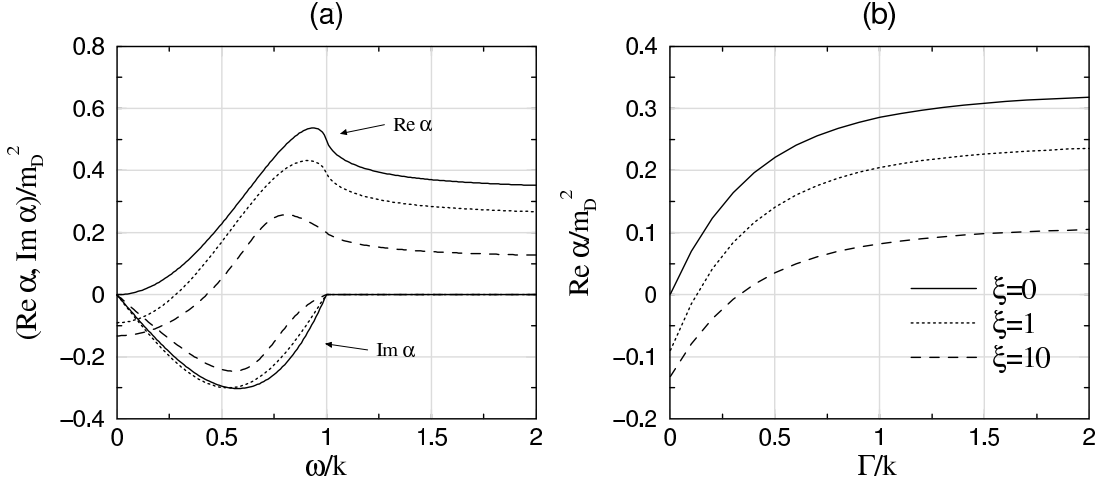}
\caption[a]{Real and imaginary parts of $\alpha/m_D^2$ as a function of real $\omega/k$ 
are shown in (a) and in (b) the real part of $\alpha/m_D^2$ is shown for $\omega/k = i \Gamma/k$ with $\theta_n = \pi/4$ and 
$\xi = \{0,1,10\}$ in both cases.}
\label{struct-fig}
\end{figure}
\bqa
\alpha(K,0) &=& \Pi_T(K) \; , \nonumber \\
\beta(K,0) &=& {\omega^2\over k^2} \Pi_L(K) \; , \nonumber \\
\gamma(K,0) &=& 0 \; , \nonumber \\
\delta(K,0) &=& 0 \; ,
\label{isolimit}
\eqa
\checked{mp}
with
\bqa
\Pi_T(K) &=& {m_D^2\over2} {\omega^2 \over k^2} 
  \left[1-{\omega^2-k^2 \over 2 \omega k}\log{\omega+k\over\omega-k}\right] \, , \\
\Pi_L(K) &=& m_D^2 \left[ {\omega\over 2k} \log{\omega+k\over\omega-k}-1\right] \; . 
\eqa
\checked{mp}
For finite $\xi$ the analytic structure of the structure functions is the same as in
the isotropic case.  There is a cut in the complex $\omega$ plane which we can chose
to run along the real $\omega$ axis from $-k<\omega<k$.  For real-valued $\omega$ the 
structure functions are complex for all $\omega < k$ and real for $\omega >k$.  For 
imaginary-valued $\omega$ all four structure functions are real-valued.
In Fig.~\ref{struct-fig} we plot the structure function $\alpha$ for real and imaginary
values of $\omega$, $\xi=\{0,1,10\}$, and $\theta_n =\pi/4$.

With these structure functions in hand we can construct the propagator $\Delta^{ij}(K)$
using the expressions from the previous section.  Writing ${\bf \Delta}^{-1}(K)$ in terms
of our tensor basis
\beq
{\bf \Delta}^{-1}(K) = (k^2 - \omega^2 + \alpha){\bf A} + (\beta - \omega^2){\bf B} 
	+ \gamma {\bf C} + \delta {\bf D} \; \,
\eeq
\checked{mp}
and applying the inversion formula (\ref{inversion}) we obtain an expression for the propagator
\beq
{\bf \Delta}(K) =  \Delta_A {\bf A} + (k^2 - \omega^2 + \alpha + \gamma)\Delta_G {\bf B}
              + [(\beta-\omega^2)\Delta_G - \Delta_A] {\bf C} - \delta \Delta_G {\bf D} \; ,
\eeq
\checked{mp}
with
\bqa
\Delta_A^{-1}(K) &=& k^2 - \omega^2 + \alpha \; , \label{propfnc1} \\
\Delta_G^{-1}(K) &=& (k^2 - \omega^2 + \alpha + \gamma)(\beta-\omega^2)-k^2 \tilde n^2 \delta^2 \; .
\label{propfnc2}
\eqa
\checked{mp}
Note that we can reorganize $\Delta(K)$ and write it as
\beq
\Delta(K) =  \Delta_A \, [{\bf A}-{\bf C}] 
	+ \Delta_G \, [(k^2 - \omega^2 + \alpha + \gamma) {\bf B} + (\beta-\omega^2) {\bf C}  - \delta {\bf D}] \; .
\eeq
\checked{mp}

\section{Static Limit}
\label{static-sec}

In order to see how the momentum-space anisotropy in the distribution functions 
affects the response to static electric and magnetic fluctuations we examine the 
limit $\omega\rightarrow0$ of the propagators (\ref{propfnc1}) and (\ref{propfnc2}).  
Approaching along the real $\omega$ axis we find that to leading order 
$\alpha \sim \gamma \sim O(\omega^0)$, $\beta \sim O(\omega^2)$, and 
$\delta \sim O(i \omega).$\footnote{Identical results can be obtained by coming
in along the imaginary axis with a suitable redefinition of $m_\delta^2$.}  
We can therefore define four mass scales
\bqa
m_\alpha^2 &=& \lim_{\omega\rightarrow0} \alpha \; , \nonumber \\
m_\beta^2 &=& \lim_{\omega\rightarrow0} - {k^2\over\omega^2} \beta \; , \nonumber \\
m_\gamma^2 &=& \lim_{\omega\rightarrow0} \gamma \; , \nonumber \\
m_\delta^2 &=& \lim_{\omega\rightarrow0}  {\tilde n k^2\over\omega} {\rm Im}\,\delta \; .
\label{massdef}
\eqa
\checked{mp}
Writing the static limit of the propagators (\ref{propfnc1}) and (\ref{propfnc2}) 
in terms of these masses gives
\bqa
\Delta_A^{-1} &=& k^2 + m_\alpha^2 \; \, \\
\Delta_G^{-1} &=& -{\omega^2\over k^2}\left[(k^2+m_\alpha^2+m_\gamma^2)(k^2+m_\beta^2)
  - m_\delta^4 \right] \; .
\eqa
\checked{mp}
$\Delta_G^{-1}$ can be factorized into 
\beq
\Delta_G^{-1} = -{\omega^2\over k^2}(k^2 + m_+^2)(k^2+m_-^2) \; ,
\eeq
\checked{mp}
where
\beq
2 m_{\pm}^2 = M^2 \pm \sqrt{M^4-4(m_\beta^2(m_\alpha^2+m_\gamma^2)-m_\delta^4)} \; ,
\label{mpm}
\eeq
\checked{mp}
with
\beq
M^2 = m_\alpha^2+m_\beta^2+m_\gamma^2 \; .
\eeq
\checked{mp}
In the isotropic limit, $\xi\rightarrow0$, $m_\alpha^2=m_\gamma^2=m_\delta^2=m_-^2=0$ and
$m_+^2 = m_D^2$.
\begin{figure}
\includegraphics[width=7cm]{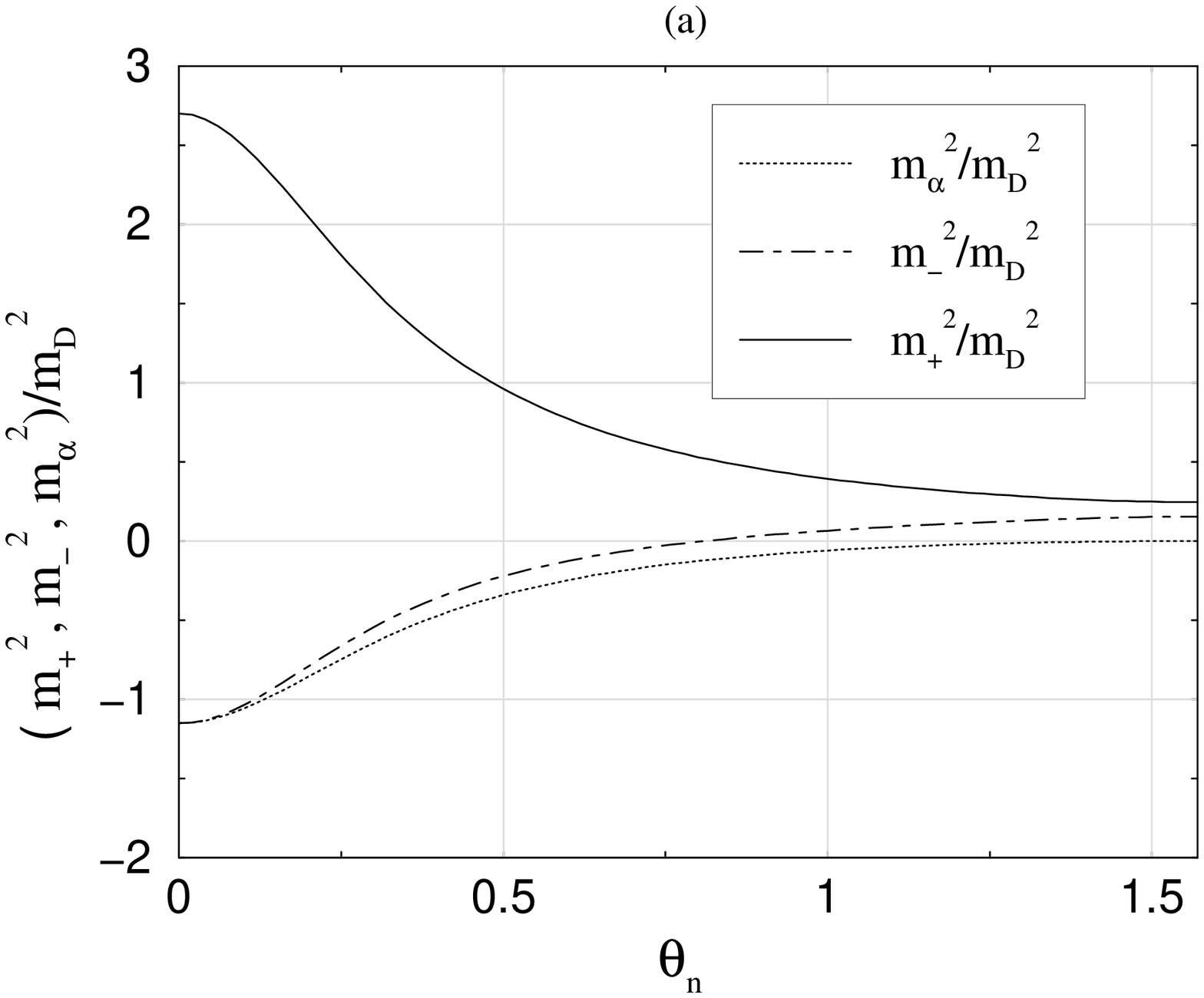}\hspace{6mm}\includegraphics[width=7cm]{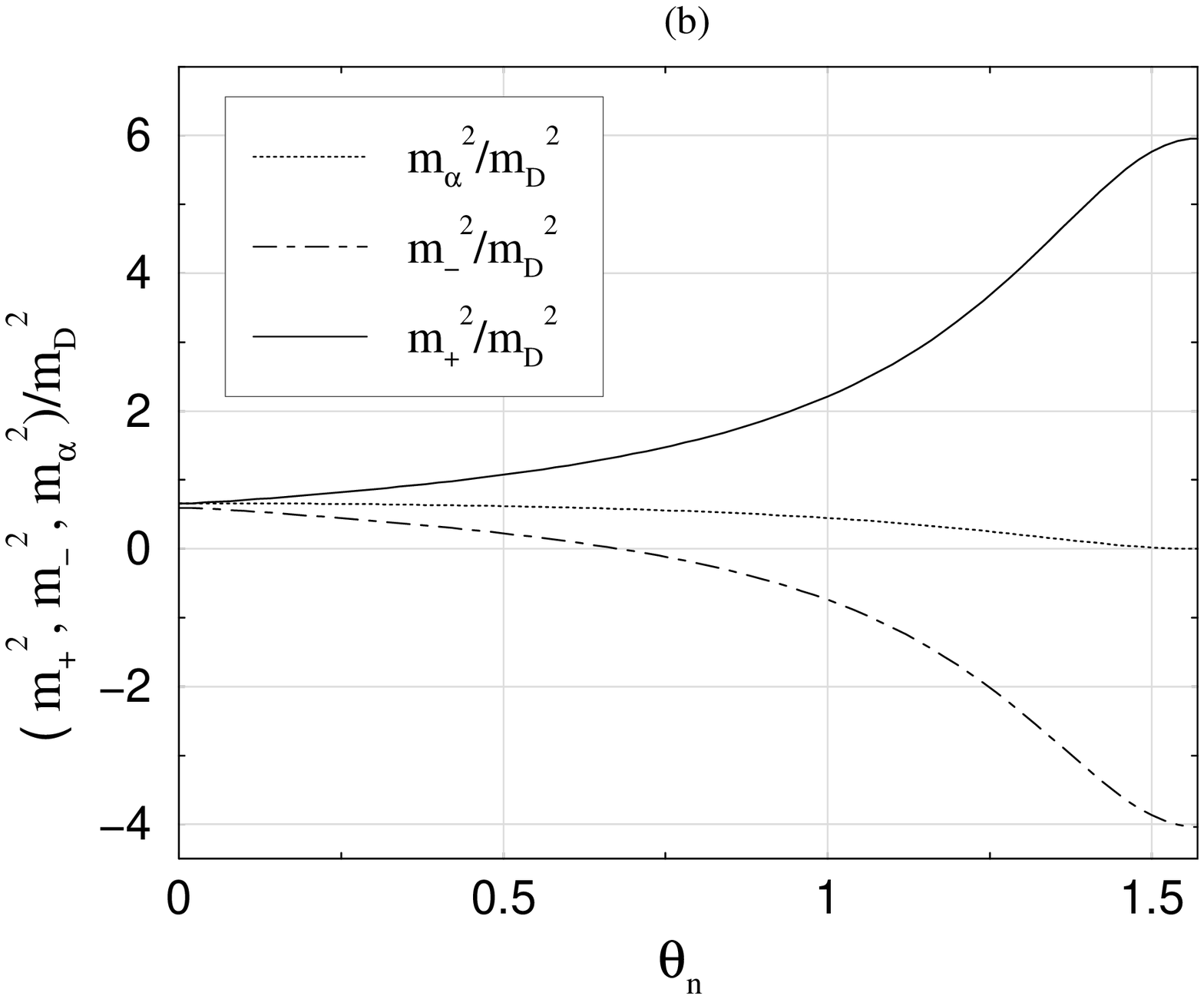}
\caption[a]{Angular dependence of $m_\alpha^2$, $m_+^2$, and $m_-^2$
at fixed (a) $\xi=10$ and (b) $\xi=-0.9$.}
\label{static-fig}
\end{figure}
For finite $\xi$ it is possible to evaluate all four masses defined above
analytically.  The results for $m_\alpha$ and $m_\beta$ are listed in
Appendix \ref{exp-app}.  In Fig.~\ref{static-fig} we plot the angular dependence of $m_\alpha^2$, $m_+^2$, 
and $m_-^2$ at fixed $\xi=10$ and $\xi=-0.9$.  
In the case $\xi>0$ (Fig.~\ref{static-fig}a) we see that for small $\theta_n$
the scale $m_+^2 \geq m_D^2$ and for $\theta_n$ near $\pi/2$, $m_+^2 \leq m_D^2$.
For small $\theta_n$ the scales $m_\alpha^2$ and $m_-^2$ are 
negative.  The fact that $m_\alpha$ and $m_-$ are non-vanishing
is in agreement with the findings of Cooper et al. \cite{CKN:2002}; however,
they neglected to consider the fact that these masses might be negative and
would therefore not correspond to screening of the magnetic interaction.
The fact that these quantities are negative indicates that for $\xi>0$ the system
possesses an instability to transverse and ``mixed'' external perturbations associated
with $m_\alpha^2$ and $m_-^2$, respectively.  The
transverse instability is present for any $\theta_n\neq\pi/2$ while the mixed
instability is only present for $\theta_n < \theta^{\rm mixed}_c$ with
$\theta^{\rm mixed}_c$ depending on the value of $\xi$.  
In the case $\xi<0$ (Fig.~\ref{static-fig}b) we see that for small $\theta_n$
the scale $m_+^2 \leq m_D^2$ and for $\theta_n$ near $\pi/2$, $m_+^2 \geq m_D^2$.
For $\theta_n \agt \pi/4$ the scale $m_-^2$ is negative again signaling the 
presence of an instability in the system.  In the next section we will discuss these instabilities 
in more detail.

\section{Collective modes}
\label{mode-sec}

A similar factorization of $\Delta_G^{-1}$ can be achieved in the non-static
case allowing us to determine the dispersion relations for all of the collective
modes in the system.  

\subsection{Stable Modes}

First, let's consider the stable collective modes which have poles at
real-valued $\omega>k$.  In this case we factorize $\Delta_G^{-1}$ as
\beq
\Delta_G^{-1} = (\omega^2 - \Omega_+^2)(\omega^2-\Omega_-^2) \; ,
\eeq
\checked{mp}
where
\beq
2 \Omega_{\pm}^2 = \bar\Omega^2 \pm \sqrt{\bar\Omega^4- 4 ((\alpha+\gamma+k^2)\beta-k^2\tilde n^2\delta^2) } \; ,
\label{omegapm}
\eeq
\checked{mp}
and
\beq
\bar\Omega^2 = \alpha+\beta+\gamma+k^2 \; .
\eeq
\checked{mp}
Note that the quantity under the square root in (\ref{omegapm}) can be written as
$(\alpha-\beta+\gamma+k^2)^2+4k^2\tilde n^2\delta^2$ which is always positive for
real $\omega>k$.  Therefore there are at most two stable modes coming from $\Delta_G$.

\begin{figure}
\includegraphics[width=14.5cm]{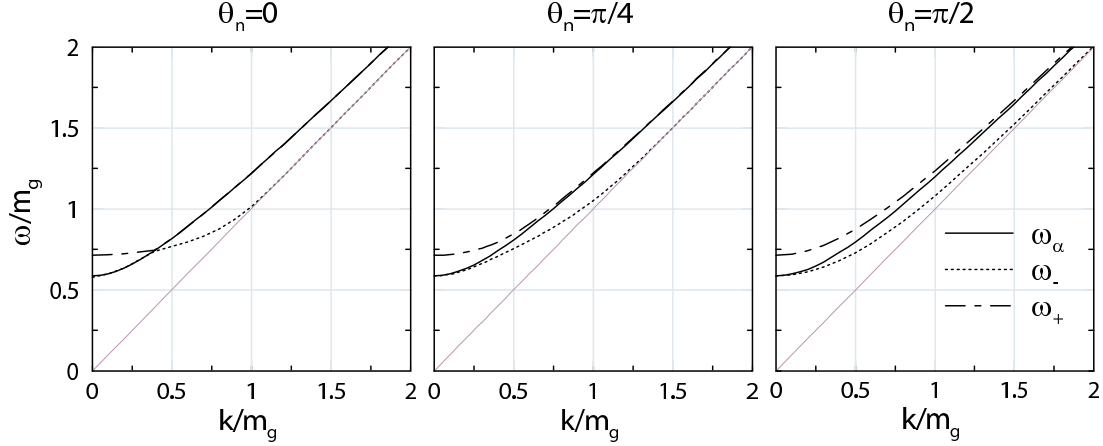}
\caption[a]{Angular dependence of $\omega_\alpha$, $\omega_+$, and $\omega_-$ for
$m_g = m_D/\sqrt{3}$, $\xi=10$, and $\theta_n=\{0,\pi/4,\pi/2\}$.}
\label{smodes-fig}
\end{figure}

The remaining stable collective mode comes from the zero of $\Delta_A^{-1}$.  The dispersion
relations for all of the collective modes can be determined by finding the solutions
to
\bqa
\omega^2_\pm &=& \Omega_\pm^2(\omega_\pm) \; , \\
\omega^2_\alpha &=& k^2 + \alpha(\omega_\alpha) \; .
\eqa
\checked{mp}
In the isotropic limit (\ref{isolimit}) $\omega_\alpha = \omega_+ = \omega_T$ 
and $\omega_- = \omega_L$.
For finite $\xi$ there are three stable quasiparticle modes with dispersion
relations which depend on the angle of propagation with respect to the anisotropy
vector, $\theta_n$.  In Fig.~\ref{smodes-fig} we plot the dispersion 
relations for all three modes for $m_D=\sqrt{3}$, $\xi=10$, and $\theta_n=\{0,\pi/4,\pi/2\}$.

\subsection{Unstable Modes}

For non-zero $\xi$ the propagator also has poles along the imaginary $\omega$ 
axis.\footnote{We have checked for poles at complex $\omega$ numerically but 
found none.}  The dispersion relation for these modes can be determined by taking
$\omega \rightarrow i \Gamma$ with $\Gamma$ real-valued and solving for 
$\Gamma(k)$.  In this case we factorize the inverse propagator as
\beq
\Delta_G^{-1} = (\Gamma^2 + \Omega_+^2)(\Gamma^2 + \Omega_-^2) \; ,
\eeq
\checked{m}
where $\Omega_\pm$ on the right hand side are evaluated at $\omega=i\Gamma$.  However,
in contrast to the stable modes there is at most one solution in this case since
numerically we find that $\Omega_+^2>0$ for all $\Gamma>0$.  

\begin{figure}
\includegraphics[height=6cm]{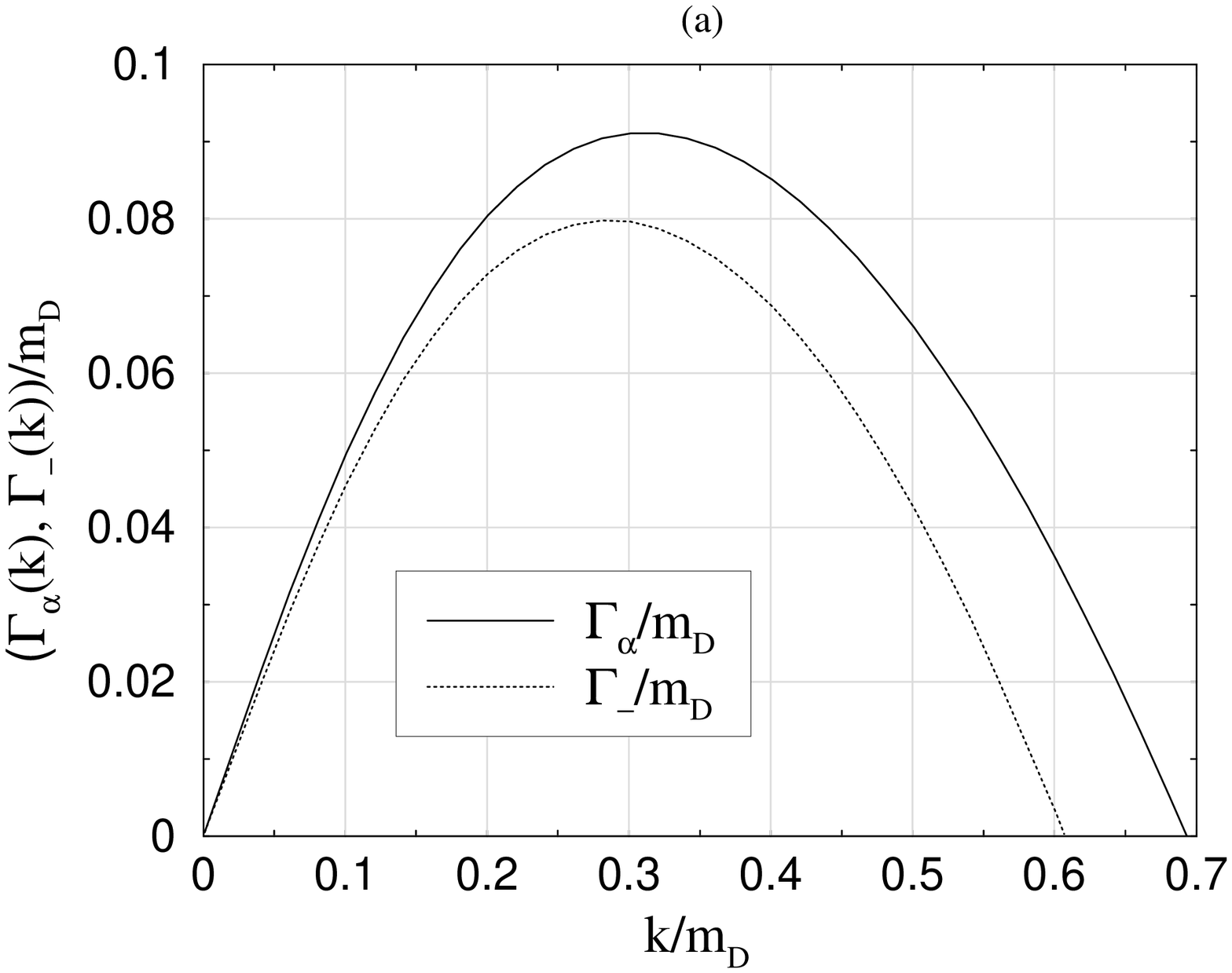}\hspace{5mm}\includegraphics[height=6cm]{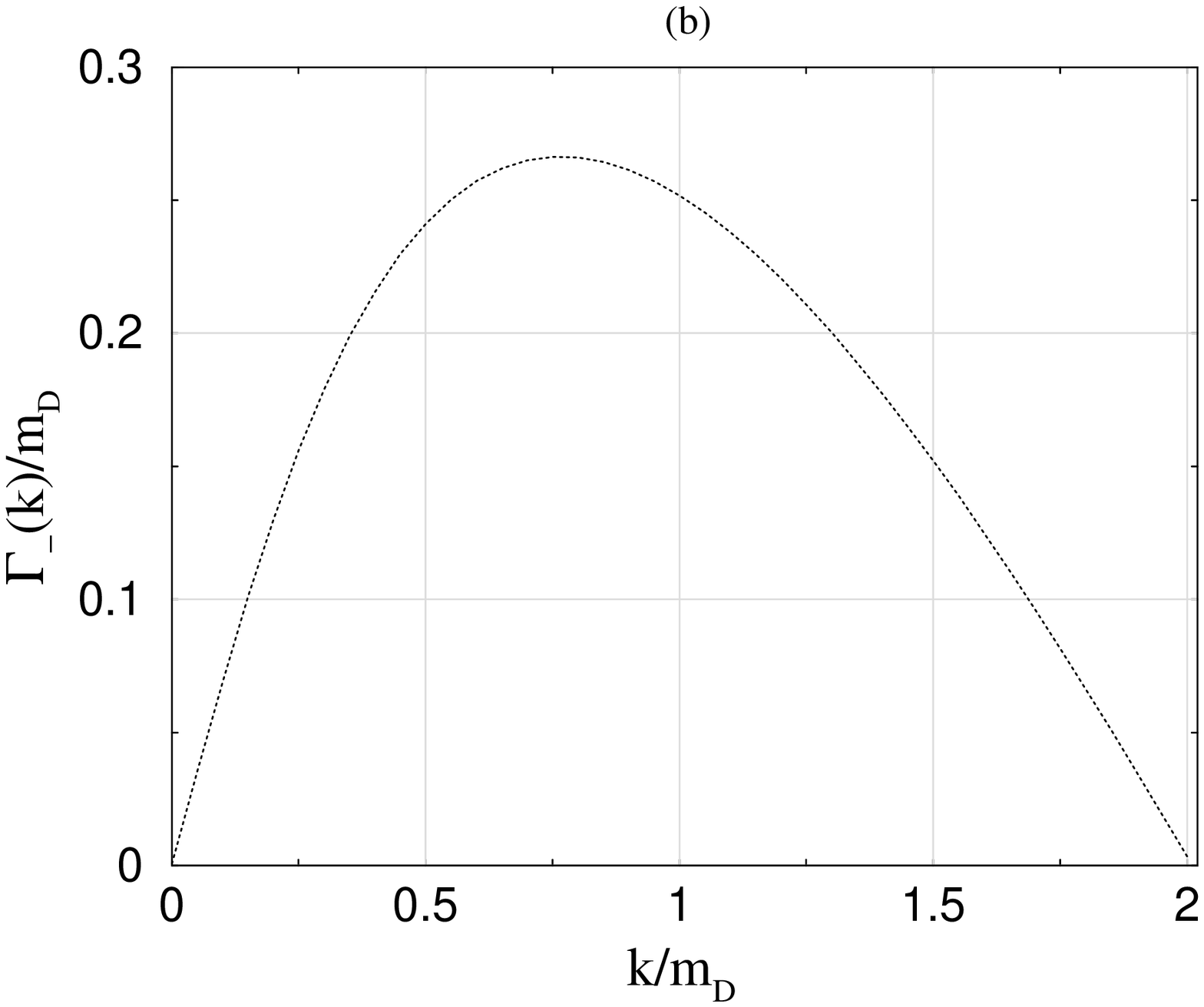}
\caption[a]{$\Gamma_\alpha(k)$ and $\Gamma_-(k)$ as a function of $k$ with
(a) $\xi=10$ and $\theta_n=\pi/8$ and (b) $\xi=-0.9$ and $\theta_n=\pi/2$.}
\label{umodes-fig}
\end{figure}

For $\xi>0$ there is also an unstable mode present in $\Delta_A$ so that in
this case there are two unstable modes in the system which can be found by solving 
\bqa
\Gamma^2_- &=& -\Omega_-^2(i \Gamma_-) \; , \\
\Gamma^2_\alpha &=& -k^2 - \alpha(i\Gamma_\alpha) \; .
\eqa
\checked{m}
Note that in both cases there are two solutions corresponding to modes 
with positive and negative growth rates.  One of these corresponds to an exponentially
growing solution and the other an exponentially decaying one.
In Fig.~\ref{umodes-fig}a we plot $\Gamma_\alpha(k)$ and $\Gamma_-(k)$ 
with $\xi=10$ and $\theta_n=\pi/8$.
For $\xi<0$ there is no longer an unstable mode coming from $\Delta_A$ and there
is therefore only one unstable mode coming from $\Gamma_-$. 
In Fig.~\ref{umodes-fig}b we plot $\Gamma_-(k)$ with $\xi=-0.9$ and $\theta_n=\pi/2$.

\section{Small $\xi$ expansion}
\label{smallxi-sec}

In the small-$\xi$ limit it is possible to obtain analytic expressions for all of the
structure functions order-by-order in $\xi$.  To linear order in $\xi$
\bqa
\alpha &=& \Pi_T(z) + \xi\Bigg[
          {z^2\over12}(3+5\cos2\theta_n)m_D^2 
          -{1\over6}(1+\cos2\theta_n)m_D^2 \nonumber \\
	  && \hspace{4cm} 
	+ {1\over4}\Pi_T(z)\left( (1+3\cos2\theta_n) - z^2(3+5\cos2\theta_n) \right)\Bigg] \; , \nonumber \\
z^{-2} \beta &=& \Pi_L(z) + \xi\left[{1\over6}(1+3\cos2\theta_n) m_D^2
	+ \Pi_L(z) \left(\cos2\theta_n-{z^2\over2}(1+3\cos2\theta_n)\right)\right] \; , \nonumber \\
\gamma &=& {\xi\over3}(3 \Pi_T(z) - m_D^2)(z^2-1)\sin^2\theta_n \; , \nonumber \\
\delta &=& {\xi\over3k}(4 z^2 m_D^2+3 \Pi_T(z)(1-4z^2))\cos\theta_n \; ,
\label{smallxi}
\eqa
\checked{mp}
where $z=\omega/k$.

\subsection{Static Limit}

Using the linear expansions and the fact that in the static limit $\Pi_L \rightarrow -m_D^2$
and $\Pi_T \rightarrow - i \pi \omega/(4 k)$ we can write for the masses (\ref{massdef}) 
\bqa
\hat m_\alpha^2 &=& - {\xi\over6}(1 + \cos 2\theta_n) \; , \nonumber \\
\hat m_\beta^2 &=& 1 + {\xi\over6}(3 \cos 2\theta_n-1) \; , \nonumber \\
\hat m_\gamma^2 &=& {\xi\over3}\sin^2\theta_n \; ,  \nonumber \\
\hat m_\delta^2 &=& -\xi{\pi\over4}\sin\theta_n\cos\theta_n \; ,
\eqa
\checked{m}
where $\hat m^2 = m^2/m_D^2$.  Using these we can obtain small-$\xi$ expressions for 
$m_\pm$ defined in (\ref{mpm})
\bqa
\hat m_+^2 &=& 1 + {\xi\over6}(3 \cos 2\theta_n-1) \; , \nonumber \\
\hat m_-^2 &=& -{\xi\over3}\cos2\theta_n \; .
\eqa
\checked{m}

\subsection{Collective Modes}

Since $\delta$ is $O(\xi)$ it can be ignored in the expansion of (\ref{omegapm}) so
that to linear order in $\xi$ the collective modes satisfy
\bqa
\Delta_A^{-1} &=& k^2-\omega^2+\alpha = 0 \nonumber \\
\Delta_G^{-1} &=& (k^2-\omega^2+\alpha+\gamma)(\beta-\omega^2) = 0 \; ,
\eqa
\checked{m}
where $\alpha$, $\beta$, and $\gamma$ are given by (\ref{smallxi}).
Note again that there is only one unstable mode coming from $\Delta_G^{-1}$ 
since $\beta(i\Gamma)>0$ for all $\Gamma>0$.

\section{Conclusions}
\label{conc-sec}

In this paper we have derived a tensor basis for the gluon self-energy in a high-temperature 
quark gluon plasma with an anisotropic momentum-space distribution.  We then restricted 
the distribution function by requiring that it could be obtained from an isotropic distribution 
function by the rescaling of one direction specified by an anisotropy vector, ${\bf \hat n}$,
and strength, $\xi$.  
Positive values of $\xi$ correspond to a contraction of the isotropic distribution function
along ${\bf \hat n}$ while negative values of $\xi$ correspond to a stretching 
along ${\bf \hat n}$.
Within this framework we could derive analytic forms for all of the structure functions
associated with the tensor basis.  Using these expressions we were then able to identify
and determine the dispersion relations for the collective modes for both positive and
negative $\xi$.  
We found
that for $\xi>0$ there were at most three stable and two unstable modes with dispersion relations 
which depended on the angle between the wave vector, ${\bf k}$, and the anisotropy 
vector.  
For $\xi<0$ we found that there were also three stable modes but only one unstable mode.
Additionally, we obtained analytic expressions for the structure
functions in the limit of small $\xi$.  These results should provide a reference 
point for the systematic study of the isotropization of a relativistic plasma.  

The study of Mr\'owczy\'nski and Randrup suggests that during heavy-ion collisions the rate 
of isotropization via collective modes is comparable with collisions and therefore cannot be 
ignored \cite{RM:2003}.  In this paper we have made no attempt to discuss the phenomenological 
rate for instability growth because there are a number of questions which would need to be 
addressed prior to making
any definitive statements about the role of instabilities in plasma evolution and their 
expected contribution to observables.  This is because we have only derived the self-energy
in a linear expansion in the fluctuations and to leading-order in the coupling constant.
Assuming that there is truly exponential growth of the fields in the direction of the 
anisotropy this means that the linear approximation will break down very quickly.  In 
practice the non-linear terms in the transport equations will become important and
regulate the growth of the modes which have become unstable.  

Within electrodynamics the coupling constant is small and it is possible to experimentally 
study the Weibel instability \cite{CLF:2002}.  
However, with QCD the story is dramatically different since for experimentally
realizable situations the coupling constant is large and the non-linear effects due to 
gluon self-interaction become important much sooner than any non-linear effects would for QED.
Nevertheless, this does not diminish from the fact that these unstable modes exist and
will therefore have a role to play in plasma evolution.  In order to assess this role,
however, detailed studies of the time evolution of anisotropic quark-gluon plasmas will need to
be performed.

\section*{Acknowledgments}
M.S. and P.R. would like to thank S.~Mr\'owczy\'nski and A.~Rebhan for discussions.  
M.S. was supported by an Austrian Science Fund (FWF) Lise Meitner fellowship M689.
P.R. was supported by the Austrian Science Fund Project No. P14632.

\appendix
\renewcommand{\theequation}{\thesection.\arabic{equation}}
\section{Notation and Conventions}
\label{not-app}

We summarize here the notation and conventions which we use in the main body of the text.

\begin{itemize}
\item{Natural units: $\hbar=c=1$}
\item{Metric: $g^{\mu\nu} = g_{\mu\nu} = {\rm diag}(1,-1,-1,-1)$}
\item{4-vectors: Indicated by Greek indices, e.g. $K^{\mu}=(\omega,{\bf k})$}.
\item{3-vectors: Indicated by lowercase Latin characters.  Upper Latin indices like $i,j,k$ 
      use a Euclidean 3-metric, e.g.  ${\bf k}=k^{i}$, $k^{i}k^{i}={\bf k}^2$}.
\item{Fourier transform:
\begin{eqnarray*}
j^{\mu}(K) & = & \int d^4 X e^{i K \cdot X} j^{\mu}(X) \\
j^{\mu}(X) & = & \int \frac{d^4 K}{(2\pi)^4} e^{- i K \cdot X} j^{\mu}(K)
\end{eqnarray*} }
\end{itemize}

\section{Analytic expressions for structure functions}
\label{exp-app}

In this appendix we collect the integral and analytic expressions for the structure functions 
$\alpha$, $\beta$, $\gamma$, and $\delta$.  Choosing ${\bf n} = \hat{\bf z}$ and ${\bf k}$ to lie 
in the $x\!\!-\!\!z$ plane ($k_x/k_z = \tan\theta_n$) we have ${\bf v}\cdot{\bf n}=\cos{\theta}$ 
and ${\bf v}\cdot{\bf k}=k_{x} \cos{\phi} \sin{\theta}+k_{z} 
\cos{\theta}$.  Using this parameterization the $\phi$ integration in all four structure
functions defined by the contractions in Eq.~(\ref{contractions}) can be performed analytically.  
\begin{eqnarray}
\alpha(K,\xi)&=&\frac{m_{D}^2}{k^2 \tilde{n}^2} \int \frac{d (\cos{\theta})}{2}%
\frac{\omega+ \xi k_{z} \cos{\theta}}{(1+\xi (\cos{\theta})^2)^2}%
\Bigg[\omega-k_{z} \cos{\theta} \nonumber\\
&& \hspace{3.5cm} +k^2 (s^2-(\cos{\theta}%
-\frac{\omega k_{z}}{k^2})^2)R(\omega-k_{z} \cos{\theta} %
,k_{x} \sin{\theta})\Bigg] ,\\
\beta(K,\xi) &=& -\frac{m_{D}^2 \omega^2}{k^2} \!\! \int \frac{d (\cos{\theta})}{2} %
\frac{1}{(1\!+\!\xi(\cos{\theta})^2)^2} \left[1\!-\!
(\omega+ \xi k_{z} \cos{\theta}) R(\omega\!-\!k_{z} \cos{\theta}, %
k_{x}\sin{\theta})\right] ,\\
\gamma(K,\xi)&=& m_{D}^2 \int \frac{d (\cos{\theta})}{2}%
\frac{1}{k^2(1+\xi \cos^2{\theta})^2}\left[\omega^2+\xi k^2 \cos^2{\theta}
-2\frac{k^2}{k_{x}^2}(\omega^2%
-\xi k_{z}^2 \cos^2{\theta})\right.\nonumber\\
&& \hspace{13mm} \left. + \frac{(\omega+\xi k_{z}\cos{\theta} )k^4}{k_{x}^2}%
\left(2(\cos{\theta}-\frac{\omega k_{z}}{k^2})^2-s^2\right)
R(\omega-k_z \cos{\theta},k_{x}\sin{\theta}) \right] , \\
\delta(K,\xi)&=&\frac{m_{D}^2 \omega}{k^4 \tilde{n}^2} \!\! \int \frac{d(\cos{\theta})}{2}%
 \frac{\omega+\xi k_{z} \cos{\theta}}{(1+\xi \cos^2{\theta})^2}%
\left(k_{z}\!+\!(k^2 \cos{\theta}\!-\!\omega k_{z}) R(\omega\!-\!k_{z} \cos{\theta},k_{x}%
\sin{\theta})\right) ,
\end{eqnarray}
\checked{m}
where $s^2=(1-\omega^2/k^2)(k_x^2/k^2)$ and
\begin{equation}
R(a,b) = \int_{0}^{2\pi} \frac{d\phi}{2\pi} \frac{1}{a
-b \cos{\phi}+i\epsilon}= \frac{1}{\sqrt{a+b+i\epsilon}\sqrt{a-b+i\epsilon}} \; .
\end{equation}
When $a$ and $b$ are real-valued $R$ can be simplified to
\begin{equation}
R(a,b) = \frac{\rm{sgn}(a) \Theta(a^2-b^2)}{\sqrt{a^2-b^2}}-
\frac{i \Theta(b^2-a^2)}{\sqrt{b^2-a^2}} \; ,
\end{equation}
\checked{m}
with $\Theta(x)$ being the usual step-function.  Note that the remaining integration over 
$\theta$ can also be done analytically but the results are rather unwieldy so we do not list them here.

\section*{Static Limit}

In the limit $\omega\rightarrow 0$ it is possible to obtain analytic expressions 
for all four structure functions.  The results for $m_\alpha$ and $m_\beta$ defined
in Eq.~(\ref{massdef}) are
\begin{eqnarray}
m_\alpha^2&=&-\frac{m_D^2}{2 k_{x}^2 \sqrt{\xi}}%
\left(k_z^2 \rm{arctan}{\sqrt{\xi}}-\frac{k_{z} k^2}{\sqrt{k^2+\xi k_{x}^2}}%
\rm{arctan}\frac{\sqrt{\xi} k_{z}}{\sqrt{k^2+\xi k_{x}^2}}\right) \; , \\
m_\beta^2&=&m_{D}^2 
\frac{(\sqrt{\xi}+(1+\xi)\rm{arctan}{\sqrt{\xi}})(k^2+\xi k_x^2)+\xi k_z\left(%
k_z \sqrt{\xi} + \frac{k^2(1+\xi)}{\sqrt{k^2+\xi k_{x}^2}} %
\rm{arctan}\frac{\sqrt{\xi} k_{z}}{\sqrt{k^2+\xi k_{x}^2}}\right)}{%
2  \sqrt{\xi} (1+\xi) (k^2+ \xi k_x^2)} \, ,
\end{eqnarray}
with similar results for $m_\gamma^2$ and $m_\delta^2$.

\bibliography{bsample}
\bibliographystyle{utphys}

\end{document}